\documentclass[%
aip,
jcp,%
amsmath,amssymb,
reprint%
,longbibliography,
]{revtex4-2}

\usepackage{amsmath}
\usepackage{amsfonts}
\usepackage{amssymb,bm}
\usepackage{graphicx}
\usepackage{physics}
\usepackage{upgreek}
\usepackage[outercaption]{sidecap}   
\usepackage{color}
\usepackage{hyperref}
\hypersetup{colorlinks=true,
linkcolor=magenta,
allcolors=magenta}
\usepackage{newtxtext}
\usepackage{newtxmath}
\usepackage[version=4]{mhchem}

\DeclareMathAlphabet{\pazocal}{OMS}{zplm}{m}{n}
\newcommand{\pP}{\ensuremath{\pazocal{P}}}

\newcommand{\pL}{\ensuremath{\pazocal{L}}}
\newcommand{\pK}{\ensuremath{\pazocal{K}}}
\newcommand{\pR}{\ensuremath{\pazocal{R}}}

\newcommand{\sys}{\ensuremath{\mathrm{s}}}

\newcommand{\bath}{\ensuremath{\mathrm{b}}}

\newcommand{\sing}{\ensuremath{\mathrm{S}}}

\renewcommand{\op}[1]{\ensuremath{\hat{#1}}}

\newcommand{\dla}{\langle\!\langle}
\newcommand{\dra}{\rangle\!\rangle}

\newcommand{\highlight}[1]{{\color{black}#1}}
\newcommand{\highlightnew}[1]{{\color{black}#1}}

\usepackage{mathtools}
\newcommand\numberthis{\addtocounter{equation}{1}\tag{\theequation}}

\begin{document}
\title{\highlight{A simple improved low temperature correction for the hierarchical equations of motion}}
\author{Thomas P. Fay\looseness=-1}
\email{tom.patrick.fay@gmail.com}
\affiliation{Department of Chemistry, University of California, Berkeley, CA 94720, USA\looseness=-1}


\begin{abstract}
	The study of  open system quantum dynamics has been transformed by the hierarchical equations of motion (HEOM) method, which gives the exact dynamics for a system coupled to a harmonic bath at arbitrary temperature and system-bath coupling strength. However in its standard form the method is only consistent with the weak-coupling quantum master equation at all temperatures when many auxiliary density operators are included in the hierarchy, even when low temperature corrections are included. Here we propose a new \highlight{low temperature} correction scheme for the termination of the hierarchy \highlight{based on Zwanzig projection} which alleviates this problem, and restores consistency with the weak-coupling master equation with a minimal hierarchy. \highlight{The utility of the new correction scheme is demonstrated on a range of model systems, including the Fenna-Metthews-Olson complex.} The new closure is found to improve convergence of the HEOM even beyond the weak-coupling limit and is very straightforward to implement in existing HEOM codes.
\end{abstract}

\maketitle

\section{Introduction}

The hierarchical equations of motion (HEOM) method, originally derived by Tanimura and Kubo,\cite{Tanimura1989,Tanimura2020} is a powerful tool for studying the dynamics of open quantum systems, where a sub-system of interest is coupled to a harmonic oscillator bath. The method has found applications in a wide range of problems, including the study of photosynthetic light-harvesting,\cite{Ishizaki2009a,Kreisbeck2014,Sarovar2010,Tong2020} electron transport,\cite{Diamond2016,Ke2022} spin relaxation,\cite{Takahashi2020} non-equilibrium heat transport,\cite{Kato2016,Hsieh2019} non-linear spectroscopy,\cite{Tanimura1989,Chen2010,Dijkstra2017} and molecular spin transport,\cite{Fay2018,Fay2021d} to name but a few examples. The method can treat problems at arbitrary temperatures and system-bath interaction strengths, and fully accounts for quantum mechanical and non-Markovian effects, and therefore it has also been an essential tool for testing approximate quantum dynamics methods.\cite{Berkelbach2012a,Fetherolf2017,Lawrence2019,Pfalzgraff2019,Mannouch2022}

The HEOM method 
 gives the exact system dynamics in terms of a formally infinite hierarchy of coupled auxiliary density operators, which has to be terminated at some finite number of terms for practical calculations. However for systems at low temperatures, the terminated HEOM solution can become unstable, leading to incorrect and sometimes highly unphysical dynamics.\cite{Dunn2019} The ``brute force'' solution to this problem is increasing the number of modes included in the hierarchy, but for systems with many baths this can dramatically increase the computational cost of HEOM calculations. One widely employed solution to correct \highlight{for modes excluded from the hierarchy} is the Ishizaki-Tanimura low-temperature correction.\cite{Ishizaki2005a} However, even with this correction, costly calculations with large hierarchies can be required in systems with very weak system-bath coupling, as will be demonstrated below.
 
\highlight{Several methods have been proposed to resolve issues of instabilities and slow convergence of the HEOM in a computationally efficient manner. The simplest approaches involve using different truncation schemes for the hierarchy,\cite{Ishizaki2005a,Hartle2013,Dijkstra2017} wherein the hierarchy is pruned according to some criterion which selects only the elements of the hierarchy which make significant contributions to the dynamics. A more sophisticated method involves removing small auxiliary density operators ``on the fly'' during an HEOM calculation to generate an optimally truncated hierarchy.\cite{Shi2009a} Convergence of the HEOM can also be improved using various alternative decompositions of the bath correlation function, such as the Pad\'e decomposition,\cite{Hu2011} and the more recently proposed Fano spectrum\cite{Cui2019} and Prony fitting schemes,\cite{Chen2022} as well as the generalized HEOM approach.\cite{Ikeda2020}
	
Other methods to improve convergence of the HEOM include modified termination schemes for the hierarchy such as the partial-ordering prescription (POP),\cite{Xu2005} Caldeira-Legget (CL) master equation truncation,\cite{Kong2015} and the time derivative truncation (TD).\cite{Hou2015} These schemes have all been demonstrated to improve convergence of the HEOM, but they come at the cost of adding complexity to the simple structure of the HEOM. For example the POP scheme adds time-dependent terms to the generator for the HEOM, and the CL and TD schemes add many couplings between auxiliary density operators in the deepest layer of the hierarchy.\cite{Kong2015,Hou2015} These methods reduce the depth of hierarchy needed to achieve a desired accuracy, but they do not directly reduce the number of modes (e.g. Matsubara frequencies) needed for convergence. A Prony filtration method has also recently been proposed to remove instabilities from the HEOM,\cite{Dunn2019} but this requires storage of the hierarchy at previous times, which could become prohibitive for large systems coupled to many baths. Collective bath coordinate mapping approaches have also been used to obtain converged HEOM dynamics in systems with strong system-bath coupling where the conventional HEOM breaks down,\cite{Ikeda2022,Li2022} but these involve an explicit treatment of a collective coordinate for each bath, which may limit these methods to systems with a small number of baths.

Another successful solution to converge the HEOM for difficult problems is to use the matrix product state (MPS)/tensor train (TT)\cite{Shi2018,Borrelli2019,Yan2021,Ke2022} as an \textit{ansatz} for the full HEOM, which enables a ``brute force'' solution to the convergence problem by facilitating calculations with much larger hierarchies. Using time-dependent variational principles these \textit{ans\"atze} can be used to find very accurate solutions to the HEOM with very large hierarchies, thereby resolving convergence issues. These methods can however be very complicated to implement efficiently and their use is likely excessive in systems with weak system bath coupling but large system energy scales, where weak-coupling quantum master equations already fairly accurate. 
 
Many of the aforementioned methods still employ the Ishizaki-Tanimura low temperature correction in some form (if any low temperature correction is used in calculations). Here we argue that the origin of slow convergence of the HEOM at low temperatures in some cases, even when system-bath coupling is weak, is in the inconsistency between the Ishizaki-Tanimura low temperature correction and the correct weak-coupling quantum master equation. In this paper a very simple solution to this problem is found by deriving a new HEOM low temperature correction, which restores consistency between the HEOM and the correct weak-coupling quantum master equation at all temperatures, using Zwanzig projection operator techniques. This new correction can easily be used with many of the advanced HEOM methods outlined above. We also demonstrate how this idea can be applied to derive an approximate terminator for the HEOM, which is closely related to the previously mentioned time-derivative truncation scheme.\cite{Hou2015} This terminator can also help to remove instabilities from the HEOM, at no additional cost in propagating the hierarchy of ADOs. We also test the new low temperature correction scheme and terminator on a set of model problems, including a model of $\sing_1 \to \sing_0$ internal conversion and a model of exciton dynamics in the Fenna-Metthews-Olson (FMO) protein-pigment complex.}

\section{Theory}
\subsection{The hierarchical equations of motion}
The HEOM method gives the exact system dynamics for a system coupled to a harmonic bath, 
\begin{align}\label{sb-ham-eq}
\op{H}  = \op{H}_\sys + \op{H}_{\bath} + \op{V} \op{B},
\end{align}
in which $\op{H}_\sys$\footnote{\highlightnew{Reorganization terms may or may not be included in $\op{H}_\sys$. In all calculations performed here we use the system Hamiltonian exactly as stated, with no additional reorganization terms added.} } and $\op{H}_\bath$ are the bare system and harmonic bath Hamiltonians, $\op{V}$ is a system operator, and the bath coupling operator $\op{B} = \sum_\alpha c_\alpha \op{q}_\alpha$ which depends linearly on the harmonic oscillator displacements $\op{q}_\alpha$.For this model, the system dynamics at a given temperature are determined fully by the spectral density of the bath $\mathcal{J}(\omega) = (\pi/2)\sum_{\alpha}(c_\alpha^2 / (m_\alpha \omega_{\alpha}))\delta(\omega - \omega_{\alpha})$, where $m_\alpha$ and $\omega_{\alpha}$ are the oscillator masses and frequencies respectively. 

With the HEOM method, the exact dynamics are obtained from a hierarchy of auxiliary system density operators (ADOs), which can be derived by differentiating the exact path-integral expression for the system density operator.\cite{Tanimura1989} The ADOs are denoted $\op{\rho}_{\vb{n}}(t)$, indexed by $\vb{n} = (n_0, n_1, \dots)$, and the physical system reduced density operator is obtained as the zeroth element of this hierarchy, i.e. $\op{\rho}(t) =\Tr_\bath[\op{\rho}_\mathrm{tot}(t)] = \op{\rho}_{\vb{0}}(t)$, assuming $\op{\rho}_\mathrm{tot}(0) = \op{\rho}(0)e^{-\beta \op{H}_\bath}/Z_\bath$.\cite{Tanimura2020} These ADOs obey the following set of coupled differential equations,
\begin{align}\label{heom-eq}
	\begin{split}
		\dv{t}\op{\rho}_{\vb{n}}(t) &= -\frac{i}{\hbar}\left[\op{H}_\sys, \op{\rho}_{\vb{n}}(t)\right] - \sum_{k} n_{k}\nu_{k}\op{\rho}_{\vb{n}}(t) \\ 
		&+\sum_{k}\left(\sqrt{n_k+1}\pL_{k-} \op{\rho}_{\vb{n}_{k}^+} + \sqrt{n_k}\pL_{k+} \op{\rho}_{\vb{n}_{k}^-}\right)
	\end{split}
\end{align}
where $\vb{n}_k^\pm = (n_0,\dots,n_k\pm1,\dots )$, $\pL_{k-} = -i\sqrt{|a_k|} [\op{V},\ \cdot\ ]$, and $\pL_{k+} = (a_k \op{V}\ \cdot\ -\bar{a}_k^*\ \cdot\ \op{V})/\sqrt{|a_k|}$. The parameters $a_k$, $\bar{a}_k$ and $\nu_k$ are obtained from the decomposition of the bath correlation function $C(t) = \Tr_\bath[\op{B}(t)\op{B}e^{-\beta \op{H}_\bath}]/Z_\bath$ as $C(t) = \sum_{k=0}^{\infty} a_{k} e^{-\nu_{k} t}\ \text{and}\ 
C(t)^* = \sum_{k=0}^{\infty} \bar{a}_{k}^* e^{-\nu_{k} t}$,\cite{Tanimura2020} where we assume $\nu_k$ are indexed in ascending order by their real part. Note that we have written the HEOM in their scaled form.\cite{Shi2009a} 

\subsection{HEOM termination}

The hierarchy in Eq.~\eqref{heom-eq} is formally infinite, so the set of auxiliary density operators has to be truncated to some finite set for practical calculations. One part of this truncation is done by only explicitly including some set of modes up to $k=M = k_\mathrm{c}-1$, in the calculation and by truncating the remaining infinite set of ADOs to some finite set. The truncation of the hierarchy for $k\leq  M$ is typically performed using various truncation criteria, such as by truncating ADOs above some maximum excitation level $L = \sum_{k=0}^M n_k$,\cite{Ishizaki2005a} or by truncating ADOs above some maximum frequency $|\gamma_{\vb{n}}| = |\sum_{k=0}^M n_k \nu_k|$.\cite{Dijkstra2017} The HEOM dynamics can then be converged to arbitrary precision by increasing the number of modes, $k$, included and the number of ADOs used for these explicitly treated modes. 

A standard approach to correct for the truncated terms in the series expansion of $C(t)$ is to treat the terms with $k\geq k_\mathrm{c}$ with a one-sided delta function, $e^{-\nu_{k} t} \approx (1/\nu_k)\delta_+(t)$, in the path integral expression for the reduced density operator.\cite{Ishizaki2005a,Tanimura2020} This gives the following correction, which is added to the right-hand side of Eq.~\eqref{heom-eq} for the explicitly treated ADOs,
\begin{align}
	\Xi_0 \op{\rho}_{\vb{n}}(t) 
	&=\sum_{k=k_\mathrm{c}}^{\infty} \nu_k^{-1}\pazocal{L}_{k-}\pazocal{L}_{k+}\op{\rho}_{\vb{n}}(t) .\label{orig-lowT-eq}
\end{align}
This type of correction is commonly referred to as a ``low temperature'' correction because it was first used to correct for truncated Matsubara frequencies terms in the Matsubara expansion of $C(t)$, which only become important at low temperatures.\cite{Ishizaki2005a,Tanimura2020}

For this approximation to be valid, we require that the approximated terms decay much faster than the characteristic system frequencies. In other words $\nu_k \gg \omega_{\sys}$ should be satisfied for $k\geq k_\mathrm{c}$. 
For large system energy scales this approximation can break down, leading to slow convergence of the HEOM dynamics with respect to $k_\mathrm{c}$, even though such systems can be well approximated with simple perturbative theories. This is because the $k_\mathrm{c}=0$ limit of $\Xi_0$ is not consistent with weak-coupling perturbative theories, as explained in Appendix \ref{wcqme-app}. Examples of this slow convergence of the HEOM are shown in the left hand panels of Figs.~\ref{fig1} and \ref{fig2} for a spin boson model (the details of which are given below).

\subsection{The Nakajima-Zwanzig Low Temperature Correction}\label{lowT-corr-sec}

An alternative modified low temperature correction term can be derived by taking a Zwanzig-projected quantum master equation perspective on the hierarchy truncation. First we note that we can write down the complete hierarchy of ADOs as
\begin{align}
	|\rho(t)\dra = \sum_{\mathrm{all}\ \vb{n}} |\rho_{\vb{n}} (t)\dra \otimes |\vb{n}\dra
\end{align}
where $|\vb{n}\dra$ is a basis vector corresponding to auxiliary density operator $\vb{n}$, and $|\rho_{\vb{n}} (t)\dra$ is the Liouville space vector of this ADO. We can write down the equation of motion for the ADOs more compactly as\cite{Shi2009b,Ikeda2022,Ke2022}
\begin{align}
	\dv{t}|\rho(t)\dra &= \pL |\rho(t)\dra = (\pL_\sys\otimes \pazocal{I}_{\mathrm{ado}} - \pazocal{I}_\sys\otimes \Gamma  + \pazocal{V}) |\rho(t)\dra
\end{align}
where $\pL_\sys = -(i/\hbar)[\op{H}_\sys,\ \cdot\ ]$ is the system Liouvillian, $\pazocal{I}_\sys$ and $\pazocal{I}_{\mathrm{ado}}$ are identity operators on the system Liouville space and the set of ADOs respectively, $\Gamma$ is diagonal matrix of decay rates for each ADO,
\highlight{\begin{align}
	\Gamma = \sum_{\mathrm{all }\ \vb{n} } \gamma_{\vb{n}}|\vb{n}\dra\dla\vb{n}|
\end{align}
in which $\gamma_{\vb{n}} = \sum_{k=0}^M n_k\nu_k$,} and $\pazocal{V}$ is the term that couples different ADOs within the hierarchy
\highlight{\begin{align}
		\begin{split}
	\pazocal{V} &= \sum_{\mathrm{all }\ \vb{n} } \sum_{k=0}^\infty\bigg(\sqrt{n_{k}+1}\pL_{k-}\otimes |\vb{n}\dra\dla\vb{n}_{k}^+| \\
	&+ \sqrt{n_{k}}\pL_{k+}\otimes |\vb{n}\dra\dla\vb{n}_{k}^-|\bigg).
	\end{split}
\end{align} }

We can now obtain an equation for the hierarchy of ADOs where the number of modes is truncated at $k=M = k_\mathrm{c}-1$, by projecting this equation with the projection operator 
$
	\pP =\sum_{\vb{n}\in\pazocal{N}_M}\pP_{\vb{n}} =  \sum_{\vb{n}\in\pazocal{N}_M}\pazocal{I}_\sys \otimes |\vb{n}\dra\!\dla\vb{n}|
$ 
where $\pazocal{N}_M$ is the set of all ADO indices where $n_k = 0$ for $k>M$.  The projected hierarchy of ADOs $\pP |\rho(t)\dra$ contains all ADOs which are treated explicitly, and its dynamics are given exactly by the Nakajima-Zwanzig equation,\cite{Nakajima1958,Zwanzig1960,Mori1965} \highlightnew{(a brief derivation of which is given in Appendix \ref{nz-app})}
\highlightnew{\begin{align}
	\begin{split}
	\dv{t}\pP|\rho(t)\dra &= \pP\pL \pP |\rho(t)\dra \\
	+  \int_0^{t} &\dd{\tau} \pP \pazocal{V}e^{(1-\pP)\pL \tau}(1-\pP)\pazocal{V} \pP |\rho(t-\tau)\dra.\label{nz-eq}
	\end{split}
\end{align}}
We can approximate the last term in this equation to second order in the coupling term $\pazocal{V}$, and apply the Markovian approximation to arrive at a perturbative form of the above equation
	\highlight{\begin{align}
	\dv{t}\pP|\rho(t)\dra &= \pP\pL \pP |\rho(t)\dra + \pK \pP |\rho(t)\dra \\
\pK &= \int_0^\infty \dd{t}\pP \pazocal{V} (1-\pP) e^{\pL_0t} (1-\pP)\pazocal{V}\pP \label{K-markov-eq} \\
	&= \sum_{\vb{n}\in\pazocal{N}_M} \Xi_{\vb{n}} \pP_{\vb{n}},
\end{align}}
where $\pL_0 = \pL_\sys\otimes \pazocal{I}_{\mathrm{ado}} - \pazocal{I}_\sys\otimes \Gamma$. \highlightnew{In this equation we have eliminated the time-convolution and replaced the upper limit of the integral with $\infty$ to make the equation time-homogeneous, an approximation which is valid when the terms in the kernel decay on a shorter time-scale than the dynamics of interest\cite{Sparpaglione1988} and for long-time dynamics.\cite{Fay2018,Fay2019}} Noting that \highlight{$(1-\pP)\pazocal{V}\pP_{\vb{n}}$} only couples ADO $\vb{n}$ to $\vb{n}_{k}^+$ for $k\geq k_\mathrm{c}$, and that \highlight{$\pP\pazocal{V}(1-\pP)$} only couples this back to $\vb{n}$, the new low temperature correction term $\Xi_{\vb{n}}$ for each ADO is found to be
\begin{align}\label{new-lowT-eq}
\Xi_{\vb{n}} = \sum_{k=k_\mathrm{c}}^\infty\pazocal{L}_{k-}\Pi_\sys (\nu_k + \gamma_{\vb{n}} - \Lambda_\sys )^{-1}\Pi_\sys^{-1}\pazocal{L}_{k+}
\end{align}
where $\Pi_\sys$ and $\Lambda_\sys$ are the matrix of eigenvectors, and eigenvalues of $\pL_\sys$, i.e. $\pL_\sys = \Pi_\sys \Lambda_\sys \Pi_\sys^{-1}$. This gives an alternative low-temperature correction which accounts for the finite system frequencies. Clearly it is very closely related to the original form in Eq.~\eqref{orig-lowT-eq}, to which it reduces if $\|\gamma_{\vb{n}}-\Lambda_\sys\|\ll \nu_k$. \highlightnew{The generalization of this correction for an arbitrary exponential approximation, for example the Pad\'e decomposition,\cite{Hu2011} is given in Appendix \ref{gencorr-app}.}

This new low temperature correction reduces the HEOM to the perturbative Markovian Nakajima-Zwanzig equation in the limit where only $\op{\rho}_{\vb{0}}(t)$ is included in the hierarchy, which in the weak-coupling limit is exact and obeys detailed balance. In this limit the original low-temperature corrected HEOM reduces to a high temperature master equation\cite{Takahashi2020} which does not satisfy detailed balance at all temperatures. This is expanded upon in the appendix. For these reasons the new low temperature corrected HEOM should converge faster in the low temperature limit. 
It is also possible to use the time-homogeneous limit of the time-convolutionless quantum master equation to close the HEOM,\cite{Breuer2001} as is outlined in the supplemental material. However we find that this approach is inferior to the Nakajima-Zwanzig approach described here, which interestingly is in agreement with previous findings that the Nakajima-Zwanzig approach gives more accurate results for systems with non-unitary reference dynamics.\cite{Fay2019b} \highlight{We also remark in passing that the Zwanzig projection approach described here could also be used to derive higher order low temperature corrections to the HEOM,\cite{Berkelbach2012a} but for now we will restrict ourselves to the lowest order correction.}

\subsection{The Diagonal Nakajima-Zwanzig Terminator}

We can also extend this projection operator treatment of the HEOM truncation to obtain a correction term for the termination of the hierarchy to some truncated set of $\vb{n} \in \pazocal{N}_{\mathrm{trunc}}$. If we only include the diagonal contributions in the hierarchy, i.e. terms that only couple $| \rho_{\vb{n}}(t) \dra$ to itself, \highlight{so $\pK$ is approximated as
\begin{align}
	\pK \approx \sum_{\vb{n} \in \pazocal{N}_{\mathrm{trunc}}}\pP_{\vb{n}} \pK \pP_{\vb{n}}.
\end{align}}
The correction term for ADO $\vb{n}$ arising from hierarchy truncation is
\begin{align}\label{trunc-eq}
	\Xi_{\vb{n}}^{\mathrm{trunc}} \!=\! \sum_{k \in {\pazocal{M}}_{\vb{n}}^{\mathrm{trunc}} } (n_k+1) \pazocal{L}_{k-} \Pi_\sys (\nu_k + \gamma_{\vb{n}} - \Lambda_\sys )^{-1} \Pi_\sys^{-1} \pazocal{L}_{k+}
\end{align}
where $\pazocal{M}_{\vb{n}}^{\mathrm{trunc}} $ is the set of mode indices at which ADO $\vb{n}$ terminates. 
Including this diagonal correction term adds no additional dense blocks to the generator ($\pP \pL \pP + \pK$) for the truncated hierarchical equations of motion with the low-temperature correction above. \highlight{Furthermore $\Xi_{\vb{n}}^{\mathrm{trunc}}$ can be pre-computed along with $\Xi_{\vb{n}}$ and used with no extra computational cost in propagating the HEOM.} 

This truncation term can be viewed as modification of Tanimura \textit{et al.}'s fast modulation scheme,\cite{Tanimura1991,Tanimura1994} and unlike \highlight{the POP scheme proposed in} Ref.~\onlinecite{Xu2005}, Eq.~\eqref{trunc-eq} does not produce a time-dependent generator in the equations of motion, which adds complexity to the time propagation. \highlight{It is interesting to note that if we applied the TCL2 approach to derive a terminator, we arrive at the POP scheme, which was originally derived from a path integral approach. Eq.~\eqref{trunc-eq} is closely related to the Caldeira-Leggett termination scheme\cite{Kong2015} and the time-derivative truncation scheme\cite{Hou2015}; indeed it is simply an approximation to the time-derivative scheme where cross couplings between $\vb{n}$ and $\vb{n}_{k,k'}^{+-} = (n_0,...,n_{k}+1,...,n_{k'}-1,...)$ are excluded.} The full perturbative correction term $\pK$ for the hierarchy truncation, \highlight{which is equivalent to the time-derivative truncation scheme}, includes these cross-terms which couple different ADOs, which adds considerable computational effort to integrating the equations of motion. We have found however that these terms can be neglected and convergence of the HEOM can still be improved with just the diagonal contributions in Eq.~\eqref{trunc-eq}. \highlight{It should be noted that this termination correction, and the termination schemes mentioned above, are separate from the low temperature corrections in Eqs.~\eqref{orig-lowT-eq} and \eqref{new-lowT-eq}, although clearly they are closely related, and here we have clarified how they can both be derived in a consistent Zwanzig projection framework.} 

\section{Numerical Tests}

\highlight{\subsection{Spin boson models}}
\begin{figure}[t]
	\includegraphics[width = 0.485\textwidth]{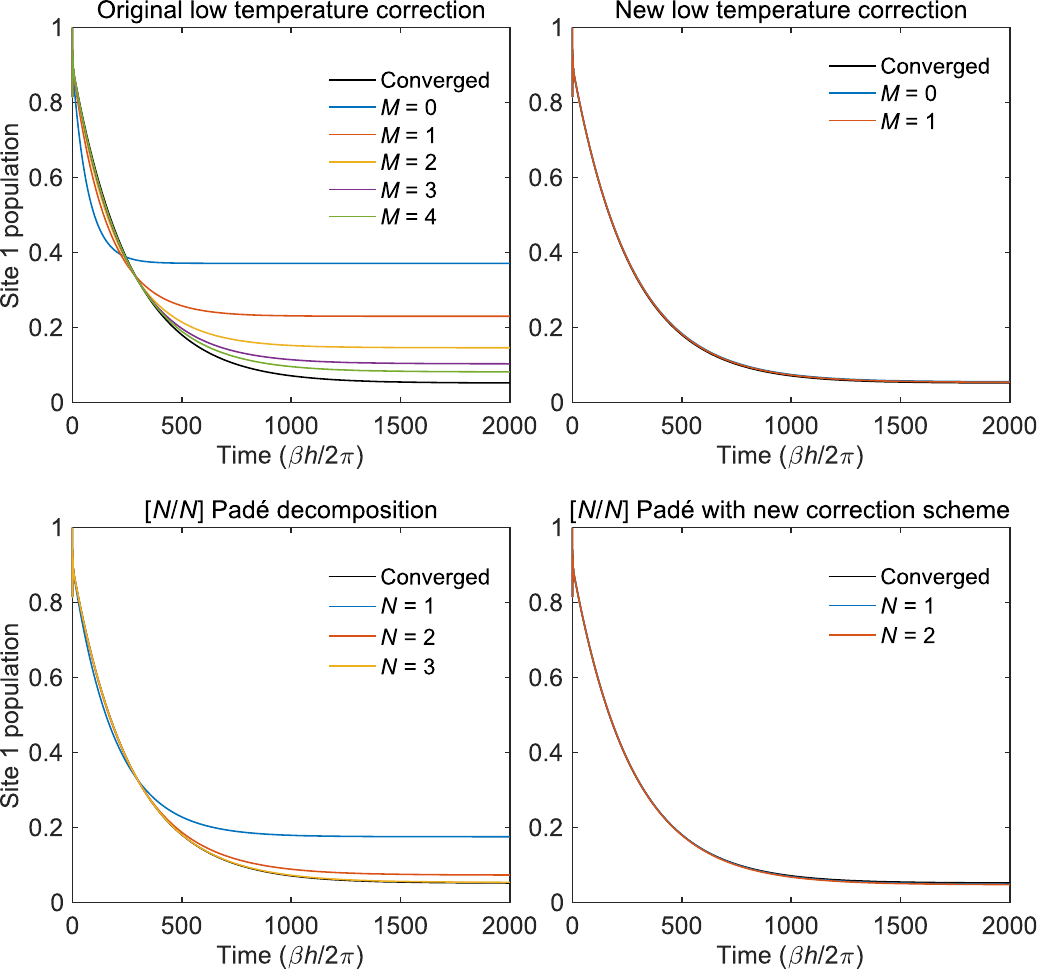}
	\caption{\highlightnew{Convergence of the site 1 population dynamics, $(1+\ev{\sigma_z(t)})/2$, for the spin boson model with $\beta\epsilon = 20$, $\beta\Delta = 5$, $\beta\hbar\omega_{\mathrm{D}} = 1$, $\beta\lambda_{\mathrm{D}} = 0.1$ with various bath correlation function decompositions and low temperature correction schemes. Converged results use $\Gamma_\mathrm{c} = 40 \omega_{\mathrm{D}}$. Top left: convergence with the original low temperature correction, top right: convergence with the new low temperature correction, bottom left: convergence with the $[N/N]$ Pad\'e decomposition, bottom right: convergence with the $[N/N]$ Pad\'e decomposition with the new low temperature correction scheme. Apart from the converged results, all calculations were performed with $L,M$ hierarchy truncation with $L=3$.}}\label{fig1}
\end{figure}

In order to test the convergence properties of the HEOM with the new low-temperature corrections, calculations were performed on the spin boson model with the two correction schemes. In this model the system Hamiltonian is given by
\begin{align}
	\op{H}_\sys = \frac{\epsilon}{2}\op{\sigma}_z + \Delta \op{\sigma}_x,
\end{align}
and $\op{V} = \op{\sigma}_z$, where $\op{\sigma}_\alpha$ are the Pauli operators, and the spectral density for the bath is taken to be a Debye spectral density 
\begin{align}\label{JD-eq}
	\pazocal{J}_{\mathrm{D}}(\omega) = \frac{\lambda_{\mathrm{D}}}{2}\frac{\omega_{\mathrm{D}} \omega }{\omega^2 + \omega_{\mathrm{D}}^2}.
\end{align}
The standard Matsubara decomposition \highlightnew{(detailed in Appendix \ref{decomp-app})} of the bath correlation function for this spectral density was used for the bath correlation function $C(t)$ to construct the hierarchy,\cite{Ishizaki2005a,Tanimura2020} in which $\nu_0 = \omega_{\mathrm{D}}$ and $\nu_{k} = 2\pi k/\hbar\beta$ for $k>0$. The hierarchy was truncated using either a frequency cut-off criterion where only ADOs with $\gamma_{\vb{n}} \leq \Gamma_\mathrm{c}$ were included,\cite{Dijkstra2017} \highlightnew{or the standard $L,M$ ``depth-based'' cut-off scheme.\cite{Ishizaki2005a}} \highlight{With the frequency-based termination scheme, $\Gamma_\mathrm{c}$ controls both $M$, with $M$ set by $M = \lfloor\hbar\beta\Gamma_\mathrm{c}/2\pi\rfloor$, and the maximum depth of the hierarchy $L$, and it also removes ADOs with $k\leq M$ and $\sum_k n_k \leq L$ with large decay rates. Stronger system-bath coupling and larger system frequencies require larger values of $\Gamma_\mathrm{c}$ to converge the HEOM dynamics. It has recently been noted that the slow $1/\omega$ decay of this spectral density at large $\omega$ can lead to slower convergence of the HEOM with respect to $M$ when compared to other, faster decaying, spectral densities,\cite{Ishizaki2020}  thus the Debye spectral density presents a challenging test for the low temperature corrections.} 

The initial condition was set to $\op{\rho}(0) = \dyad{1} = (\op{1}+\op{\sigma}_z)/2$  and other parameters are given in the figure captions. For all calculations an adaptive short iterative Arnoldi algorithm (which is described in the supplemental material) was used to integrate the equations of motion. Here we compare the standard Ishizaki-Tanimura low-temperature correction, Eq.~\eqref{orig-lowT-eq}, to using Eqs.~\eqref{new-lowT-eq} and \eqref{trunc-eq} to close the HEOM.

\begin{figure}[t]
	\includegraphics[width = 0.485\textwidth]{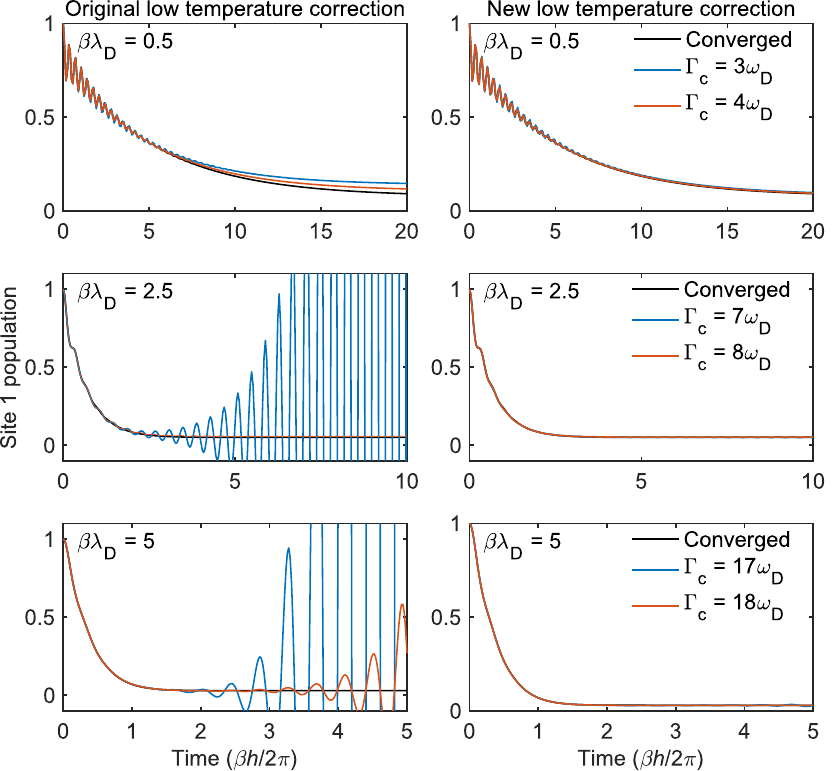}
	\highlight{\caption{Convergence of the HEOM dynamics for the spin boson model with $\beta\epsilon = 15$, $\beta\Delta = 5$, $\beta\hbar\omega_{\mathrm{D}} = 5$, $\beta\lambda_{\mathrm{D}} = 0.5$ (top), $2.5$ (middle) and $5$ (bottom), for the original low temperature correction (left) and the new low temperature correction and termination scheme (right). Converged results use $\Gamma_\mathrm{c} = 20 \omega_{\mathrm{D}}$ for $\beta\lambda=0.5$ and $\beta\lambda=2.5$, and $\Gamma_\mathrm{c} = 22 \omega_{\mathrm{D}}$ for $\beta\lambda_{\mathrm{D}} = 5$. The blue and red lines correspond to calculations with $\Gamma_\mathrm{c}/\omega_{\mathrm{D}} = 3$ and $4$ (top panels), $\Gamma_\mathrm{c}/\omega_{\mathrm{D}} = 7$ and $8$ (middle panels), and $\Gamma_\mathrm{c}/\omega_{\mathrm{D}} = 17$ and $18$ (bottom panels).}\label{fig2}}
\end{figure}
In Fig.~\ref{fig1} we see that even for a case with very weak system-bath coupling, where the reorganization energy, $\beta \lambda_{\mathrm{D}} = 0.1$, is roughly 100 times smaller than the system energy scales, the original low temperature corrected HEOM converges very slowly with respect to the cut-off parameter $M$, with contributions from Matsubara terms up to $M=3$ (when $\Gamma_\mathrm{c} = 6\pi/\beta$) still not being sufficient to converge the results. Conversely the new low-temperature correction scheme gives essentially converged results for a hierarchy with \highlightnew{$M=0$}. 

\highlightnew{We have also performed calculations for the above example with the $[N/N]$ Pad\'e decomposition scheme for the correlation function,\cite{Hu2011} where $N = M$ sets the maximum number of modes in the decomposition. Further details of this scheme can be found in Refs.~\onlinecite{Hu2011} and \onlinecite{Ding2012}. It has been suggested that the $[N/N]$ approximant should optimal for the Debye spectral density,\cite{Ding2012} so we have only considered this version of the approximant in this example. We see in the bottom left panel of Fig.~\ref{fig1} that the Pad\'e approximant also suffers from the same issues as the Matsubara decomposition with the Ishizaki-Tanimura correction when a small number of modes $N$ are included in the hierarchy, although the Pad\'e approximant does converge faster with respect to the maximum number of modes included in the hierarchy, converging by $N=3$. In the bottom right panel we show the results of calculations with the $[N/N]$ Pad\'e scheme combined the new low temperature correction scheme, where we see again the new scheme accelerates convergence. This shows that issues of convergence in low temperatures exist for both Matsubara and Pad\'e decompositions of $C(t)$, neither of which are fully consistent with the weak-coupling quantum master equation at low temperatures for finite $M$.}

\highlight{The same trend is seen in more challenging cases in Fig.~\ref{fig2}, in which $\beta \lambda_{\mathrm{D}} = 0.5,2.5,5$ and $\beta\hbar \omega_{\mathrm{D}} = 5$, where results converge faster with $\Gamma_\mathrm{c}$ with the new correction scheme, whereas the original scheme is less accurate, and is unstable at larger reorganization energies and gives unphysical negative populations with the same sizes of hierarchy. The instability in the hierarchy is particularly evident in these examples at $\beta \lambda_{\mathrm{D}} = 5$. }

\begin{figure}[t]
	\includegraphics[width=0.485\textwidth]{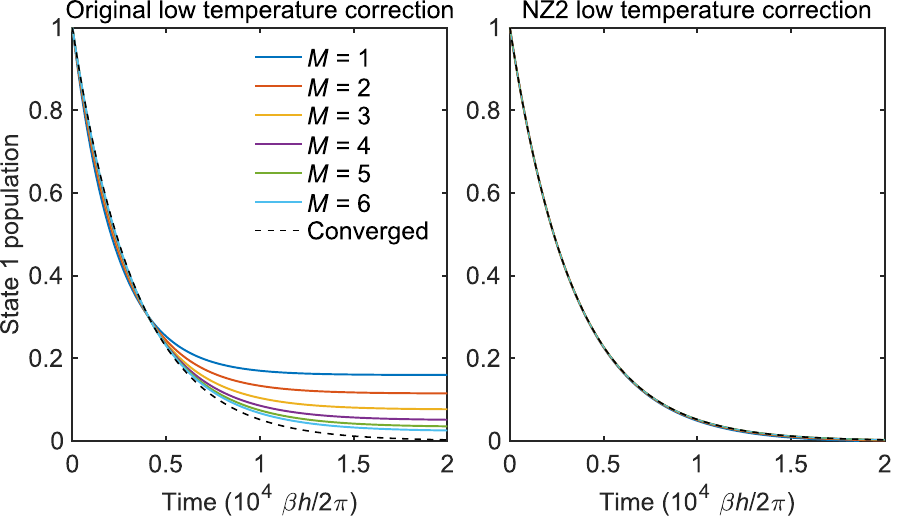}
	\caption{\highlight{Convergence of the original low temperature correction (left) and the new low temperature correction (right) with respect to $M$ for the state 1 population dynamics in the internal conversion model. Converged results use the new low temperature scheme with $L=3$ and $M=4$, whilst all other results use $L=2$.}}\label{ic-fig}
\end{figure}

\highlight{\subsection{Internal conversion model}}
\highlight{
As another demonstration of the utility of the new low temperature correction scheme, we have performed calculations on a model of the $\sing_1\to \sing_0$ internal conversion in a rigid chromophore. We model this as a system with two electronic states, $\ket{0}$ and $\ket{1}$, Hamiltonian $\op{H}_\sys = \epsilon\dyad{1}$, with $\beta\epsilon = 30$ 
, which is coupled to two harmonic baths. One bath is a tuning bath, which modulates the $\ket{0}$-$\ket{1}$ energy gap, with coupling operator $\op{V}_\mathrm{t} = \dyad{0}$. The tuning bath spectral density is taken as a Debye spectral density (Eq.~\eqref{JD-eq}), with $\beta \lambda_{\mathrm{D}} = 0.15$, and $\beta\hbar\omega_{\mathrm{D}} = 1$, which is approximately representative of the spectral density for the $\sing_0$-$\sing_1$ energy gap in bacteriochorophyll-a at 300 K.\cite{Sarovar2010} The second bath is a coupling bath, which couples $\ket{0}$ and $\ket{1}$ via the coupling operator $\op{V}_\mathrm{c} = \dyad{0}{1}+\dyad{1}{0}$. The spectral density for the coupling bath is taken to be a Brownian Oscillator spectral density,\cite{Tanimura1994}
\begin{align}
	\pazocal{J}_{\mathrm{c}}(\omega) = \frac{\lambda_{\mathrm{c}}}{2}\frac{\gamma\Omega^2\omega}{(\omega^2-\Omega^2)^2+\gamma^2\omega^2},	
\end{align}
with $\beta\hbar\Omega = 7.5$ and $\beta\hbar\gamma = 1$, which corresponds to an under-damped oscillator with a frequency of about 1650 cm${}^{-1}$ at 300 K, typical of a high frequency molecular vibrational mode, and we set $\beta \lambda_{\mathrm{c}} = 0.03$ in these calculations. \highlightnew{The Matsurbara decomposition used in this work for this spectral density is detailed in Appendix \ref{decomp-app}.} In this example we have used the $L$, $M$ truncation scheme,\cite{Ishizaki2005a} truncating the hierarchy to include all ADOs with $\sum_{j=\mathrm{t,c}}\sum_k n_{jk} \leq L$ and $k\leq M$. 

In this example we have just used the new low temperature correction, Eq.~\eqref{new-lowT-eq}, and we have not used the truncation correction, Eq.~\eqref{trunc-eq}, in order to demonstrate the utility of the new low temperature correction on its own. In Fig.~\ref{ic-fig} we show the state $\ket{1}$ population dynamics for an initial condition $\op{\rho}(0) = \dyad{1}$, for the Ishizaki-Tanimura low temperature correction and the new low temperature correction. Again we see that the new low temperature correction scheme gives results that converge much faster with respect to $M$, being essentially converged with $M=2$, whereas the Ishizaki-Tanimura scheme has not converged with $M=6$. This reflects the fact that the transition rates between states are proportional to $\Re[\tilde{C}_\mathrm{c}(\pm\epsilon/\hbar)]$ to lowest order in the system-bath coupling, where $\tilde{C}_{\mathrm{c}}(\omega)$ is the Fourier transform of the coupling bath correlation function. The new low temperature correction scheme is completely consistent with the weak-coupling master equation, so it produces rates that agree with this limit for any value of the truncation parameter $M$, and therefore it accurately predicts long time populations. The Isizaki-Tanimura scheme however does not correctly capture this feature, which manifests as incorrect long time dynamics. This example demonstrates the utility of the new low temperature correction in a system with mutiple baths, and with a spectral density other than the $\pazocal{J}_\mathrm{D}(\omega)$. It is interesting to note that the choice of low temperature correction is still important in this example, despite the faster decay of $\pazocal{J}_{\mathrm{c}}(\omega)$ at large $\omega$ ($\sim 1/\omega^3$) compared to the Debye spectral density.}
\highlight{\begin{figure}
	\includegraphics[width = 0.485\textwidth]{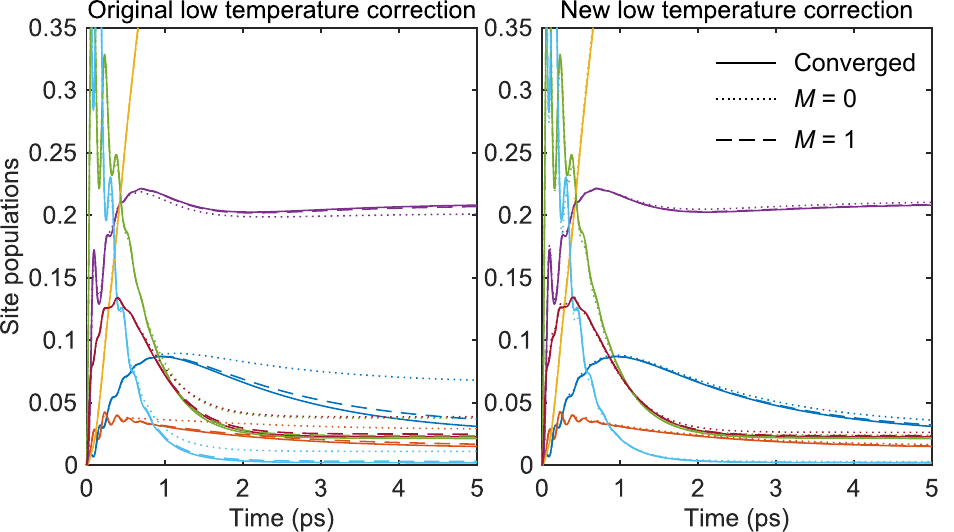}
	\caption{A comparison of the original and new low temperature corrections for the FMO site populations at 77 K with the original low $T$ correction (left), and the new scheme (right). Converged results use the new correction with $L=3$, $M=3$. The colors correspond to: blue BChla1, red BChla2, gold BChla3, purple BChla4, green BChla5, cyan BChla6, and burgundy BChla7. }\label{fig3}
\end{figure}}
\\
\highlight{
\subsection{Fenna-Metthews-Olson complex}
The improved convergence of the new low temperature correction can be very useful in systems with multiple baths, where adding modes to the hierarchy significantly increases computational cost. As an example we have simulated the excitation energy transfer in the Fenna-Metthews-Olson (FMO) complex,\cite{Engel2007} which has both stimulated significant debate since early investigations into excitation dynamics in the complex\cite{Ishizaki2009a,Sarovar2010,Cao2020,Runeson2022} and serves as an important model system for approximate quantum dynamics methods.\cite{Pfalzgraff2019,Runeson2020a,Mannouch2022} The FMO model consists of seven bacteriochorophyll-a (BChla) chromophores where local excitations on individual chromophores couple to each other, and to the surround protein environment. The environment coupling is described with a bath for each chromophore, with coupling operator $\op{V}_n = \dyad{n}$, described by a single Debye spectral density, \highlightnew{exactly as in Eq.~\eqref{JD-eq}}, with $\lambda_{\mathrm{D}} = 35\text{ cm}^{-1}$ and $\hbar\omega_{\mathrm{D}} =106.1\text{ cm}^{-1}$, and with the system Hamiltonian from Ref.~\onlinecite{Sarovar2010}. The initial system condition was set as a local excitation on BChla6, $\op{\rho}(0) = \dyad{6}$, and the temperature was set to $T= 77$ K. The $L,M$ cut-off scheme was used for all calculations with $L = 3$ and $M$ varied for each correction scheme. Converged results used the new correction scheme with $L=3$ and $M=3$.

The simulation results are shown in Fig.~\ref{fig3} where it can be seen that for the same size of hierarchy the new low temperature correction with the diagonal truncation correction yields more accurate the long time populations than the original truncation scheme, with differences noticeable by $t = 2$ ps. Specifically the populations of high energy sites are predicted almost quantitatively with $L=3$, $M=0$ (with a hierarchy of just 120 ADOs, compared to 680 for $M=1$ and 4495 for $M=3$), whereas the same hierarchy with the original low temperature correction significantly overestimates the populations of these states. These incorrect long time populations for the Ishizaki-Tanimura scheme clearly arise from breaking detailed balance in the weak coupling limit. The accelerated convergence with respect to $M$ could be particularly useful in calculating more complex observables than site populations, such as 2D electronic spectra.\cite{Engel2007,Tanimura2020} Another example illustrating the shortcomings of the Ishizaki-Tanimura low temperature correction compared to the new scheme in a multi-state system is presented in the supplemental material (Sec. S.1) for a donor-bridge-acceptor model, where the original low temperature corrected HEOM can significantly overestimate transfer to a high energy bridge state.
}
\\
\section{Concluding remarks}

In conclusion, we have shown that using perturbative Nakajima-Zwanzig theory we can obtain corrections for truncation of the HEOM which restore consistency with the weak-coupling quantum master equation \emph{at all temperatures}. In the high temperature limit this closure reduces to the widely used Ishizaki-Tanimura closure,\cite{Ishizaki2005a} but at lower temperatures it accounts for finite system frequencies unlike the Ishizaki-Tanimura correction. The new correction scheme is found to improve convergence of the HEOM in several cases, including in a model of energy transport in a photosynthetic protein-pigment complex, with negligible additional computational effort.

{The new low temperature correction scheme can be used in place of the Ishizaki-Tanimura correction along with many of the methods for improving HEOM convergence and efficiency, because many of these involve some truncation of a mode expansion of $C(t)$ at some $k=M$.\cite{Shi2009a,Yan2014,Dunn2019,Hu2011,Cui2019,Chen2022,Ikeda2022,Ikeda2020,Li2022} \highlightnew{We have presented one such generalization in Appendix \ref{gencorr-app} where we have demonstrated how to include this low temperature scheme for arbitrary exponential approximations to $C(t)$, for example the Pad\'e scheme.} For example this low temperature can easily be applied along with Prony filtration,\cite{Dunn2019}  on-the-fly truncation of the hierarchy,\cite{Shi2009a} or approaches based on collective bath coordinate mapping.\cite{Ikeda2022,Li2022}} It should also be noted that the low temperature correction and terminator presented here can be applied with any cut-off criterion for the HEOM.\cite{Dijkstra2017,Ishizaki2005a,Hartle2013} 

\highlight{The Zwanzig projection approach described here could also be used to derive more accurate closures for the HEOM. For example higher order corrections can be derived, which can be approximately resummed to infinite order,\cite{Chen2016} or it may be possible to derive self-consistent approximations to improve these closures.\cite{Sowa2020} These possibilities certainly warrant further investigation. Another possible use of the Zwanzig approach is that it yields an estimate for the truncated part of the hierarchy $(1-\pP)|\rho(t)\dra\approx -\pL_0^{-1}(1-\pP)\pazocal{V}\pP |\rho(t)\dra$. This could potentially be used with the Shi \textit{et al.}'s ``on the fly'' truncation scheme\cite{Shi2009a} to reintroduce truncated ADOs in a consistent and adaptive manner.

Finally, it should be noted that the low temperature correction scheme presented here is remarkably simple. This means it can be very easily implemented in existing HEOM codes,\cite{Kreisbeck2014,Johansson2013,Ikeda2020,Pyrho} in addition to the publicly available Matlab code ``HEOM-lab'',\cite{heom-lab}  which was used to perform the calculations presented in this paper. Given this, we anticipate this new low temperature correction could be used routinely in future HEOM calculations.}
%
%
%
%
%
%

\section*{Acknowledgements}

TPF kindly thanks David Limmer and David Manolopoulos for their comments on the manuscript. TPF was supported by the U.S. Department of Energy, Office of Science, Basic Energy Sciences, CPIMS Program Early Career Research Program under Award No. DE-FOA0002019.

\section*{Supplemental material}

The supporting information includes calculations using the different low temperature corrections for the donor-bridge acceptor model (Sec. S.1), an exploration of using Markovian time-convolutionless master equation theory (i.e. Redfield theory) instead of Nakajima-Zwanzig theory to terminate the HEOM (Sec. S.2), and a description of the adaptive short iterative Arnoldi algorithm used to propagate the equations of motion in this work (Sec. S.3).

\section*{Author declarations}

\subsection*{Conflict of interest}

The author has no conflict of interest.

\section*{Data availability}

Data presented in the paper are available from the author upon a reasonable request. Code used to perform the simulations is publicly available at \url{https://github.com/tomfay/heom-lab}.

\appendix

\highlightnew{\section{The Nakajima-Zwanzig equation}\label{nz-app}

In this appendix we briefly outline the derivation of the Nakajima-Zwanzig equation and its Markovian limit. First we note that we can split the equaiton of motion for $|\rho(t)\dra$ into equations for $\pP |\rho(t)\dra$ and $(1-\pP)|\rho(t)\dra$ as
\begin{align}
	\dv{t}\pP  |\rho(t)\dra &= \pP \pL \pP  |\rho(t)\dra + \pP \pL (1-\pP)  |\rho(t)\dra \\
	\begin{split}
	\dv{t}(1-\pP)  |\rho(t)\dra &= (1-\pP) \pL \pP  |\rho(t)\dra \\
	&+ (1-\pP) \pL (1-\pP)  |\rho(t)\dra.
	\end{split}
\end{align}
The equation for $(1-\pP)|\rho(t)\dra$ can be integrated formally to give
\begin{align}
	(1-\pP)  |\rho(t)\dra = \int_0^t \dd{\tau} e^{(1-\pP)\pL (t-\tau)} (1-\pP) \pL \pP |\rho(\tau)\dra,
\end{align}
where we have assumed $\pP |\rho(t=0)\dra = |\rho(t=0)\dra$. Inserting this into the above equation for $\pP |\rho(t)\dra$, noting that $\pP \pL_0 (1-\pP) = (1-\pP) \pL_0 \pP = 0$, and using the idempotency of $\pP$, together with a change of variables in the integral $\tau \to t-\tau$, straightforwardly gives the Nakajima-Zwanzig equation, Eq.~\ref{nz-eq}. 

The Markovian limit of this equation can be obtained by Laplace transforming the Nakajima-Zwanzig equation to turn the time-convolution into a simple product term,\cite{Sparpaglione1988}
\begin{align}
	s \pP|\tilde{\rho}(s)\dra - |\rho(0)\dra &= \pP \pL \pP |\tilde{\rho}(s)\dra + \tilde{\kappa}(s) \pP |\tilde{\rho}(s)\dra
\end{align}
where $\tilde{f}(s) = \int_0^\infty \dd{t}f(t)e^{-s t}$, and $\kappa(t) = \pP \pL e^{(1-\pP)\pL t} \pP \pL \pP$. To obtain the time-homogeneous Markovian approximation to the Nakajima-Zwanzig equation we replace $\tilde{\kappa}(s)\to\tilde{\kappa}(s=0)$, and then transform the equation back to the time-domain.\cite{Sparpaglione1988} This approximation is valid when the kernel $\kappa(t)$ decays faster than the dynamics of $\pP |\rho(t)\dra$, and it exactly preserves the long-time limit and time-integrated properties of the $\pP |\rho(t)\dra$ dynamics.\cite{Fay2019} Making a perturbative approximation to the projected propagator in $\kappa(t)$, together with this approximation, yields Eq.~\eqref{K-markov-eq}}.

\highlightnew{\section{Generalized low-temperature correction}\label{gencorr-app}

In this appendix we derive a generalized version of the low temperature correction for the HEOM based on an arbitrary exponential approximation to the exact $C(t)$. Such approximations include the Pad\'e scheme\cite{Hu2011} and the Prony fitting scheme.\cite{Chen2022}  We denote such an approximation by
\begin{align}
	\tilde{C}(t) = \sum_{k=0}^{k_\mathrm{c}-1} \tilde{c}_{k} e^{-\tilde{\gamma}_k t} \text{ and }\tilde{C}(t)^* = \sum_{k=0}^{k_\mathrm{c}-1} \tilde{\bar{c}}_{k}^* e^{-\tilde{\gamma}_k t}
\end{align}
and we assume we know an exact (possibly infinite) expansion $C(t) = \sum_{k=0}^\infty c_k e^{-\gamma_k t}$ and $C(t) = \sum_{k=0}^\infty \bar{c}_k e^{-\gamma_k t}$ (for example the full Matsubara expansion). With the generalized HEOM formalism, we can write an HEOM in terms of the exact expansion of $C(t)$ as
\begin{align}
	C(t) &= \tilde{C}(t) + \delta C(t).
\end{align}
The hierarchy is expanded as a sum of exponentials $C(t) = \sum_{k=0}^{\infty} a_{k} e^{-\nu_{k} t}\ \text{and}\ 
C(t)^* = \sum_{k=0}^{\infty} \bar{a}_{k}^* e^{-\nu_{k} t}$ with decay rates $\vb*{\nu} = (\tilde{\vb*{\gamma}},\tilde{\vb*{\gamma}},\vb*{\gamma})$, and coefficients $\vb*{a} = (\tilde{\vb*{c}},-\tilde{\vb*{c}},\vb*{c})$, $\bar{\vb*{a}} = (\tilde{\bar{\vb*{c}}},-\tilde{\bar{\vb*{c}}},\bar{\vb*{c}})$, $[\vb*{\nu}]_k = \nu_k$ and likewise for the other row vectors. Now the terms from $k=0$ to $k=k_\mathrm{c}-1$ are treated explicitly, and the remaining terms, corresponding to $\delta C(t)$, are projected out as described in Sec.~\ref{lowT-corr-sec}. This gives the following generalized low temperature correction term
\begin{align}
	\begin{split}
	\Xi_{\vb{n}} &= \sum_{k=0}^\infty{\pazocal{L}}_{k-}\Pi_\sys (\gamma_k + \gamma_{\vb{n}} - \Lambda_\sys )^{-1}\Pi_\sys^{-1}{\pazocal{L}}_{k+} \\
	&-  \sum_{k=0}^{k_{\mathrm{c}}-1} \tilde{\pazocal{L}}_{k-}\Pi_\sys (\tilde{\gamma}_k + \gamma_{\vb{n}} - \Lambda_\sys )^{-1}\Pi_\sys^{-1}\tilde{\pazocal{L}}_{k+}
	\end{split}
\end{align}
where $\pL_{k-} = -i\sqrt{|c_k|} [\op{V},\ \cdot\ ]$ and $\pL_{k+} = (c_k \op{V}\ \cdot\ -\bar{c}_k^*\ \cdot\ \op{V})/\sqrt{|c_k|}$ and $\tilde{\pL}_{k-} = -i\sqrt{|\tilde{c}_k|} [\op{V},\ \cdot\ ]$ and $\tilde{\pL}_{k+} = (\tilde{c}_k \op{V}\ \cdot\ -\tilde{\bar{c}}_k^*\ \cdot\ \op{V})/\sqrt{|\tilde{c}_k|}$. This can be extended to the Ikeda \& Scholes' generalized HEOM\cite{Ikeda2020} in the same way; all that is needed is a decomposition of $\delta C(t)$ as a sum of exponentials.}

\section{Relationship to the weak-coupling master equation}\label{wcqme-app}

In order to obtain a weak-coupling quantum master equation, we can directly apply Markovian second order Nakajima-Zwanzig theory to the full sub-system plus bath Hamiltonian, Eq.~\eqref{sb-ham-eq}, with the projection operator $\pP = \op{\rho}_\bath \Tr_\bath[\ \cdot\ ]$ to obtain a weak-coupling quantum master equation which is valid at all temperatures.\cite{Nakajima1958,Zwanzig1960,Mori1965} This equation is given by
\begin{align}\label{wc-qme-eq}
	\dv{t} \op{\rho}(t) = \pL_\sys \op{\rho}(t) + \pR \op{\rho}(t)
\end{align}
where the relaxation superoperator $\pR$ is given by
\begin{align*}\label{wc-R-eq}
	\pR \op{\rho} &= -\int_0^\infty \dd{t} \Tr_\bath[[\op{V}\op{B},e^{(\pL_\sys + \pL_{\bath}) t} [\op{V}\op{B},\op{\rho}\op{\rho}_\bath]]] \\
	&= -\int_0^\infty \dd{t} \op{V}^\times\Tr_\bath[\op{B} e^{(\pL_\sys + \pL_{\bath})  t} [\op{V}\op{B},\op{\rho}\op{\rho}_\bath]]]\\
	&= -\int_0^\infty \dd{t} \op{V}^\times e^{\pL_\sys t}\left( C(t)\op{V}\op{\rho} - C(t)^*\op{\rho}\op{V}\right) \numberthis
\end{align*}
where $\op{V}^\times = [\op{V},\ \cdot \ ]$, and $\pL_\bath = -\frac{i}{\hbar}[ \op{H}_\bath,\ \cdot \ ]$. Inserting the decomposition of $C(t)$ and $C(t) = \sum_{k=0}^\infty a_ke^{-\nu_k t}$, and $C(t)^* = \sum_{k=0}^\infty \bar{a}_k^*e^{-\nu_k t}$, and inserting the eigen-decomposition of $\pL_\sys = \Pi_\sys \Lambda_\sys \Pi_\sys^{-1}$, we obtain
\begin{align}
	\pR = \sum_{k=0}^\infty \pL_{k-} \Pi_\sys(\nu_k - \Lambda_\sys)^{-1}\Pi_\sys^{-1}\pL_{k+}
\end{align}
which is exactly $\Xi_{\vb{n}=\vb{0}}$ with $k_\mathrm{c}=0$. Within the secular approximation, when the system bath coupling is weak and coherences and populations decouple in the system energy eigenbasis $\ket{E_n}$ (i.e. when the secular approximation becomes valid), the energy eigenstate population transfer rates, $\pR_{E_n E_n\leftarrow E_m E_m} = \Tr_\sys[\dyad{E_n} (\pR \dyad{E_m})]$, satisfy detailed balance,\cite{Zwanzig2001}
\begin{align}
	\frac{\pR_{E_n E_n\leftarrow E_m E_m}}{\pR_{E_m E_m\leftarrow E_n E_n}} = e^{-\beta (E_n - E_m)}.
\end{align}

For the standard low-temperature correction, when $k_\mathrm{c} = 0$ we have\cite{Takahashi2020}
\begin{align}
	\Xi_0\op{\rho} = -\int_0^\infty \dd{t} \op{V}^\times \left( C(t)\op{V}\op{\rho} - C(t)^*\op{\rho}\op{V}\right),
\end{align}
which is not consistent with the weak-coupling master equation, Eqs.~\eqref{wc-qme-eq} and \eqref{wc-R-eq}, except in the high temperature limit. This means the standard low-temperature correction to the HEOM breaks down in the low temperature limit when $k_\mathrm{c} = 0$, even for arbitrarily weak system-bath coupling. In this limit the detailed balance condition is not satisfied, which will lead to errors in long time populations.

\highlightnew{\section{The Matsubara decomposition scheme}\label{decomp-app}

In this appendix we give the precise Matsurbara decomposition schemes used for the Debye and under-damped Brownian Oscillator baths used in this work, and implemented in HEOM-lab.\cite{heom-lab} The Matsubara decomposition scheme used for the Debye spectral density correlation function is given by,\cite{Ishizaki2005a}
\begin{align}
	\nu_0 &= \omega_{\mathrm{D}} \\
	\nu_k &= 2\pi k / (\hbar\beta), \ k>0 \\
	a_0 &= \bar{a}_0 = \lambda_{\mathrm{D}}\hbar\omega_{\mathrm{D}}\left(\coth(\frac{\hbar\omega_{\mathrm{D}}\beta}{2}-i)\right) \\
	a_k &= \bar{a}_k = \frac{4\lambda_{\mathrm{D}}\omega_{\mathrm{D}}}{\beta}\frac{\nu_k}{\nu_k^2 - \omega_{\mathrm{D}}^2}. \ k> 0.
\end{align}
For the under-damped Brownian oscillator spectral density we use the labels $k = +$ and $k=-$ for the ``zeroth'' non-Matsubara terms and $k=1,2,3,...$ for the Matsubara terms. The expressions for $\nu_k$, $a_k$ and $\bar{a}_k$ are given by\cite{Tanimura1994,Ikeda2020,Lindoy2019}
\begin{align}
	\nu_\pm &= \gamma/2 \pm i \sqrt{\Omega^2 + (\gamma/2)^2}\\
	\nu_k &= 2 \pi k / (\hbar \beta), \ k = 1,2,3,... \\
	a_\pm &= \bar{a}_{\mp} = \mp \frac{\lambda_{\mathrm{c}}\Omega^2\hbar}{2\sqrt{\Omega^2 + (\gamma/2)^2}} \left(\coth(\frac{i\hbar\nu_{\pm}\beta}{2})-1\right) \\
	a_k &=\bar{a}_k =  \frac{\lambda_{\mathrm{c}}\Omega^2\gamma}{4\beta} \frac{\nu_k}{\nu_k^2\gamma^2-(\nu_k^2 +\Omega^2)^2}, \ k = 1,2,3,... .
\end{align}}
\bibliography{bibliography.bib}
\end{document}


\title{Supplemental Material: A simple improved low temperature correction for the hierarchical equations of motion}
\author{Thomas P. Fay\looseness=-1}
\email{tom.patrick.fay@gmail.com}
\affiliation{Department of Chemistry, University of California, Berkeley, CA 94720, USA\looseness=-1}


%

\maketitle

\section{Donor-bridge-acceptor system}

As an additional example of the utility of the new Nakajima-Zwanzig here we have performed HEOM calculations on a donor-bridge-acceptor (DBA) model, which consists of three electronic states, $\ket{\mathrm{D}}$, $\ket{\mathrm{B}}$ and $\ket{\mathrm{A}}$, coupled to a single Debye bath by the coupling operator $\op{V} = \frac{1}{2}\dyad{\mathrm{B}}+\dyad{\mathrm{A}}$. The Hamiltonian of the system is taken as
\begin{align}
	\op{H}_\sys = \left(\epsilon_\mathrm{B}+\frac{\lambda_\mathrm{D}}{4}\right)\dyad{\mathrm{B}} + \left(\epsilon+{\lambda_\mathrm{D}}\right) + \Delta\left(\dyad{\mathrm{D}}{\mathrm{B}}+\dyad{\mathrm{B}}{\mathrm{D}}\right) + \Delta\left(\dyad{\mathrm{A}}{\mathrm{B}}+\dyad{\mathrm{B}}{\mathrm{A}}\right).
\end{align}
and the initial condition is taken to be $\op{\rho}(0) = \dyad{\mathrm{D}}$, corresponding to the system being at local equilibrium on the donor site. In these calculations $\beta\lambda_{\mathrm{D}} = \beta\hbar\omega_{\mathrm{D}} = 1$, $\beta\Delta = 0.5$ and $\beta \epsilon_\mathrm{B} = 20$. We consider two values of $\epsilon$, the symmetric case with $\epsilon = 0$, and the biased case where $\beta\epsilon = -2$. We perform calculations with the new and Ishizaki-Tanimura low temperature correction, using the frequency cut-off scheme with the Matsubara decomposition of the bath correlation function. Converged results are obtained with $\Gamma_\mathrm{c} = 30 \omega_{\mathrm{D}}$, which includes contributes from modes up to $M = 4$ using the new low temperature correction scheme. We present results for $\Gamma_\mathrm{c} = 10 \omega_{\mathrm{D}}$ for the different low temperature corrections. The truncation corrections were not included in these examples.
\begin{figure}[h]
	\includegraphics[width=0.5\textwidth]{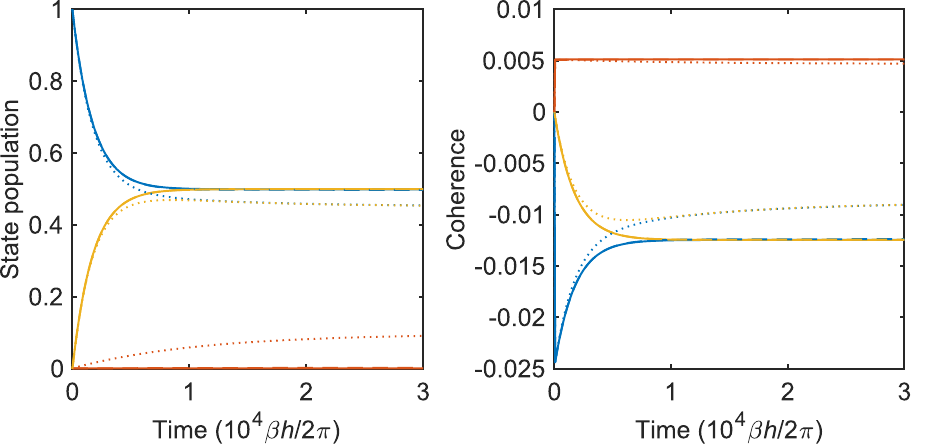}
	\caption{Populations (left) and real coherences part of the coherences (right) for the symmetric DBA model. Solid lines are the converged results, dashed lines are the new low temperature correction results and dotted lines correspond to the Ishizaki-Tanimura correction. In the left panel the lines are colored by state with D in blue, B in red and A in gold. In the right panel lines are colored by coherence with D-B in blue, D-A in red and B-A in gold.}\label{dba-symm-fig}
\end{figure}
\begin{figure}[h]
	\includegraphics[width=0.5\textwidth]{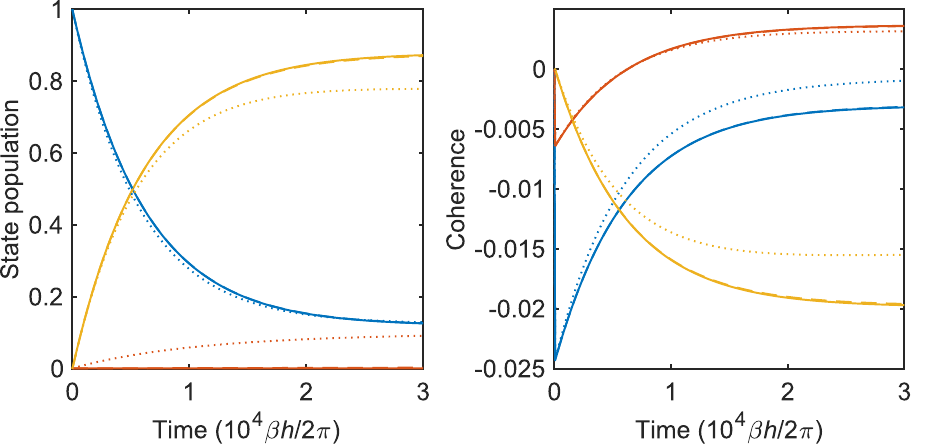}
	\caption{Same as above for the biased DBA model.}\label{dba-biased-fig}
\end{figure}

In Figs.~\ref{dba-symm-fig} and \ref{dba-biased-fig} we show the population and real-part of the coherences for the symmetric and biased DBA models. We see that for the same size of hierarchy (with modes up to $M=1$ treated explicitly) the new low temperature correction gives more accurate population and coherence dynamics than the original Ishizaki-Tanimura correction. In particular the original correction overestimates the population transfer to the bridge state quite significantly, and in the biased case it also underestimates population transfer to the A state quite significantly.

\section{The Redfield correction scheme}

Using the projection operator formalism describing in the main note, we can also obtain a Redfield correction scheme by instead using the second order time-convolutionless quantum master equation\cite{Breuer2001} in the Markovian limit with the same projection operator on the ADOs. The Redfield superator is obtained as
\begin{align}
	\pK = \int_0^\infty \dd{t}\pP \pazocal{V} e^{\pL_0 t} \pazocal{V}e^{-\pL_0 t}\pP,
\end{align}
which gives the following low temperature correction term,
\begin{align}
	\Xi_{\vb{n}} = \sum_{k=M+1}^\infty\pazocal{L}_{k-}\Pi_\sys (\mathsf{J}_{k} \circ \Pi_\sys^{-1}\pazocal{L}_{k+}\Pi_\sys) \Pi_\sys^{-1}
\end{align} 
where $\circ$ denotes the element-wise (Schur) product, and $\mathsf{J}_k$ is defined by
\begin{align}
	[\mathsf{J}_k]_{n,m} = \frac{1}{\nu_k - (\lambda_{\sys,n} - \lambda_{\sys,m})},
\end{align} 
where $\lambda_{\sys,n}$ is the $n$th eigenvalue of the system Liouvillian, $\pL_\sys = -i[\op{H}_\sys,\cdot]$. Analogously the RF truncation correction is
\begin{align}
	\Xi_{\vb{n}}^{\mathrm{trunc}} \!=\! \sum_{k \in {\pazocal{M}}_{\vb{n}}^{\mathrm{trunc}} } (n_k+1) \pazocal{L}_{k-} \Pi_\sys (\mathsf{J}_{k} \circ \Pi_\sys^{-1}\pazocal{L}_{k+}\Pi_\sys) \Pi_\sys^{-1}.
\end{align}
It should be noted that if we do not apply the Markovian approximation to the truncation term (and if we also include all cross terms), then the truncation term becomes equivalent to Xu \textit{et al.}'s POP scheme.\cite{Xu2005}

\begin{figure}
	\includegraphics[width = 0.475\textwidth]{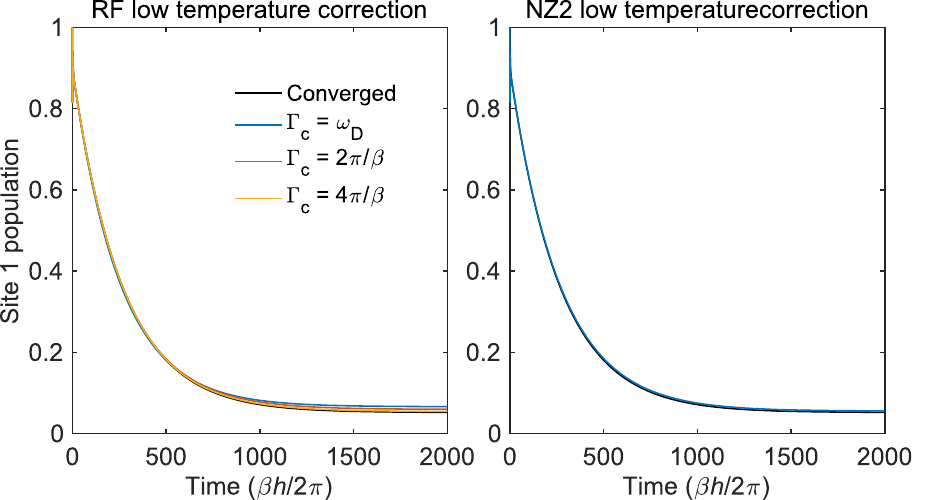}
	\caption{Convergence of the site 1 population dynamics, $(1+\ev{\sigma_z(t)})/2$, for the spin boson model with $\beta\epsilon = 20$, $\beta\Delta = 5$, $\beta\hbar\omega_{\mathrm{D}} = 2$, $\beta\lambda_{\mathrm{D}} = 0.5$, comparing the RF and NZ2 correction schemes. }\label{figs1}
\end{figure}
\begin{figure}
	\includegraphics[width = 0.475\textwidth]{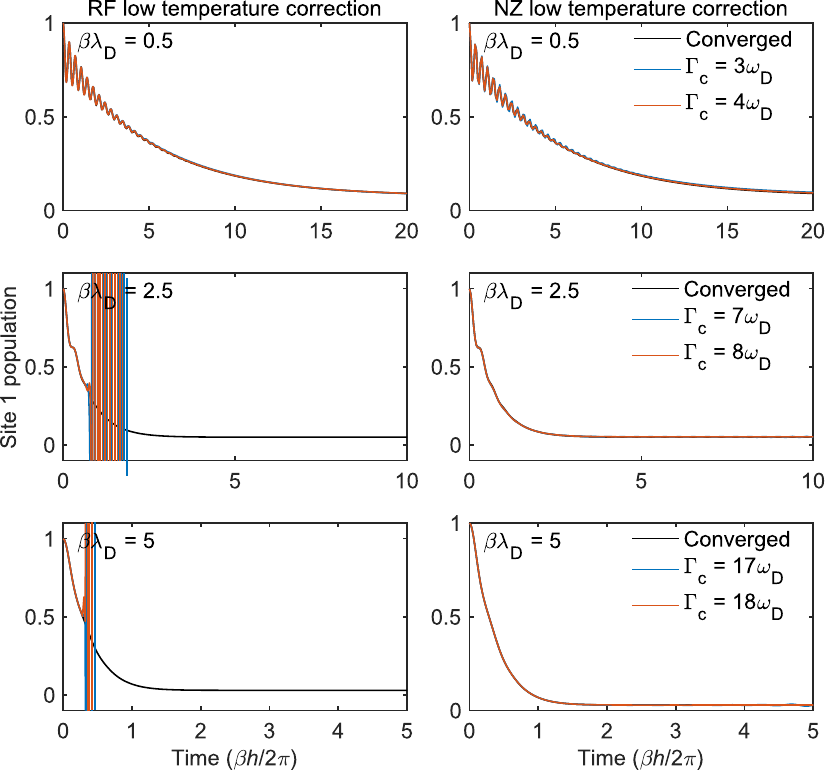}
	\caption{Convergence of the HEOM dynamics for the spin boson model with $\beta\epsilon = 15$, $\beta\Delta = 5$, $\beta\hbar\omega_{\mathrm{D}} = 5$, $\beta\lambda_{\mathrm{D}} = 0.5$ (top), $2.5$ (middle) and $5$ (bottom), for the RF low temperature correction and truncation scheme (left) and the NZ low temperature correction and termination scheme (right). Converged results use $\Gamma_\mathrm{c} = 20 \omega_{\mathrm{D}}$ for $\beta\lambda=0.5$ and $\beta\lambda=2.5$, and $\Gamma_\mathrm{c} = 22 \omega_{\mathrm{D}}$ for $\beta\lambda_{\mathrm{D}} = 5$. The blue and red lines correspond to calcualtions with $\Gamma_\mathrm{c}/\omega_{\mathrm{D}} = 3$ and $4$ (top panels), $\Gamma_\mathrm{c}/\omega_{\mathrm{D}} = 7$ and $8$ (middle panels), and $\Gamma_\mathrm{c}/\omega_{\mathrm{D}} = 17$ and $18$ (bottom panels).}\label{figs2}
\end{figure}
\begin{figure}
	\includegraphics[width = 0.475\textwidth]{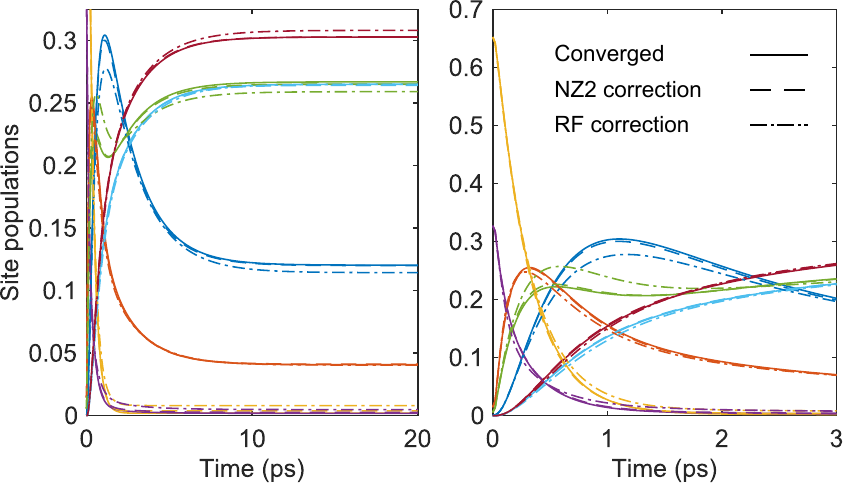}
	\caption{A comparison of the original and new low temperature corrections for the LHCII 7-mer site populations at 150K, comparing the RF and NZ2 correction schemes. Converged results use the new correction with $\Gamma_\mathrm{c} = 1980\ \text{cm}^{-1}$, which has contributions from ADOs with up to $L=5$ and $M=3$, and the other results $\Gamma_\mathrm{c} = 1320\ \text{cm}^{-1}$, which have contributions from ADOs with up to $L=3$ and $M=2$. The colours correspond to: blue 602a, red 603a, gold 608b, purple 609b, green 610a, cyan 611a, and burgundy 612a. }\label{figs3}
\end{figure}
In Figs.~\ref{figs1} and \ref{figs2} we compare the Redfield (RF) and second order Nakajima-Zwanzig (NZ2), correction schemes for the models presented in the main note. For both the low reorganisation energy spin-boson model (Fig.~\ref{figs1}) and other higher reorganisation energy cases (Fig.~\ref{figs2}) the RF scheme is less accurate and in the latter case it is unstable for $\beta\lambda_{\mathrm{D}}\geq 2.5$, and breaks down at even shorter times than the original low-temperature correction scheme without any truncation corrections. Although for the $\beta\lambda_{\mathrm{D}}= 0.5$ case in Fig.~\ref{figs2} the RF scheme does perform marginally better than the NZ scheme. Also in the LHCII 7-mer model (results in Fig.~\ref{figs1}) the RF scheme has larger errors at short and long times than the NZ2 scheme, with very little difference between the RF and original low temperature correction scheme.

\section{Adaptive short-iterative Arnoldi integration scheme}

All of the results presented here are obtained by integrating the HEOM with an adaptive short-iterative Arnoldi algorithm.\cite{Pollard1994} The HEOM is first expressed as
\begin{align}
	\dv{t}|\rho(t)\dra = \pL |\rho(t)\dra.
\end{align}
At a given $t_0$ the $k$-dimensional Krylov subspace is constructed iteratively using the Arnoldi algorithm.\cite{Pollard1994} This subspace $\mathcal{K}_k$ spans the set of vectors $  \{|\rho(t_0)\dra,\pL|\rho(t_0)\dra,\dots,\pL^{k-1}|\rho(t_0)\dra\}$, and its basis is denoted $\mathcal{B}_k = \{|q_j\dra| j = 0,\dots,k-1\}$. The basis is constructed by setting
\begin{align}
	|q_0\dra &= \frac{1}{\||\rho(t_0)\dra\|}|\rho(t_0)\dra \\
	|q_j\dra &= \frac{1}{\left\|\pL |q_{j-1}\dra - \sum_{l=0}^{j-1} |q_{l}\dra\dla q_l|\pL |q_{j-1}\dra\right\|}\left(\pL |q_{j-1}\dra - \sum_{l=0}^{j-1} |q_{l}\dra\dla q_l|\pL |q_{j-1}\dra\right)\text{ for }j = 1,\dots,k-1
\end{align}
Using the Arnoldi algorithm the matrix representation of $\pL$ on $\mathcal{K}_k$ in basis $\mathcal{B}_k$, denoted $\vb{L}_k$, which has upper Hessenberg form,\cite{Pollard1994} can be constructed at the same time as $\mathcal{B}_k$. The full algorithm for $j\geq 0$ for this procedure is
\begin{align*}
		|q_{j+1}\dra &\gets \pL |{q_{j}}\dra \\ 
		[\vb{L}_k]_{i,j} &\gets \dla q_i | q_{j+1}\dra\text{ for }i = 0,\dots,j\\
		| q_{j+1}\dra &\gets  | q_{j+1}\dra - \sum_{i=0}^j [\vb{L}_k]_{i,j}| q_{i}\dra \numberthis \\
		[\vb{L}_k]_{j+1,j} &\gets \| | q_{j+1}\dra \| \\
		| q_{j+1}\dra &\gets | q_{j+1}\dra /[\vb{L}_k]_{j+1,j}.  \\
\end{align*}

The low-dimensional representation of $|\rho(t)\dra$ is then propagated exactly with
\begin{align}
	\vb*{c}_k(t_0+\delta t) = \exp(\vb{L}_k \delta t) \vb*{c}_k(t_0)
\end{align}
where $c_{k,j}(t_0) = \delta_{j,0}{\||\rho(t_0)\dra\|}$, and $|\rho(t_0+\delta t)\dra \approx \sum_j c_{k,j}(t_0+\delta t)|q_j\dra$. This approximation is accurate to order $k-1$, and as $\delta t$ increases so does the error, so the subspace needs to be reconstructed iteratively to propagate the state to long times. In the adaptive version of this we employed, we simultaneously propagate $\vb*{c}_k(t)$ and $\vb*{c}_{k+1}(t)$, and recompute the Krylov subspace when
\begin{align}
	\frac{\|\vb*{c}_k(t_0+\delta t) - \vb*{c}_{k+1}(t_0+\delta t)\|}{\|\vb*{c}_k(t_0+\delta t)\|} > \epsilon
\end{align}
where $\epsilon$ is a convergence parameter (here $\vb*{c}_k(t_0+\delta t)$ is implicitly padded with a zero at the end to calculate the difference). In the calculations presented here $k=16$ and $\epsilon = 10^{-10}$ are used. 

Vector representations of observables $[\vb*{v}_O]_j = \dla q_j |O^\dag \dra$ can be computed when $\mathcal{K}_k$ is constructed and observables can be obtained as
\begin{align}
	\ev{O(t)} =\dla O^\dag | \rho(t) \dra= \vb*{v}_O^\dag \cdot \vb*{c}(t)
\end{align}
without needing to reconstruct $|\rho(t)\dra$.

\bibliography{si-bibliography.bib}